%% file: p.tex
\documentclass[a4paper, twocolumn]{article}
\usepackage{graphicx}
\usepackage{amsmath,amstext,amssymb}
\usepackage{cite}
\usepackage{color}
\newcommand{\taumodel}{$\tau$-model}
\newcommand{\adhoc}{\textit{ad hoc}}

\newcommand{\etal}{{\it et~al.}}%
\newcommand{\eg}{{\it e.g.}}%
\newcommand{\ie}{{\it i.e.}}%
\newcommand{\Eq}[1]{Eq.\,(\ref{#1})}%
\newcommand{\Eqs}[1]{Eqs.\,(\ref{#1})}%
\newcommand{\Fig}[1]{Fig.\,\ref{#1}}%
\newcommand{\Tab}[1]{Table~\ref{#1}}%
\newcommand{\Pep}{e$^+$}%
\newcommand{\Pem}{e$^-$}%
\newcommand{\Pee}{\Pep\Pem}%
\newcommand{\PZ}{\ensuremath{\mathrm{Z}}}
\newcommand{\eV}{\hbox{\ensuremath{\mathrm{e\kern-0.1em V}}}}%
\newcommand{\MeV}{\hbox{\ensuremath{\mathrm{M}}\eV}}%
\newcommand{\GeV}{\hbox{\ensuremath{\mathrm{G}}\eV}}%
\newcommand{\TeV}{\hbox{\ensuremath{\mathrm{T}}\eV}}%

\newcommand{\abs}[1]{\left|#1\right|}                        
\newcommand{\ycut}{\ensuremath{y_\mathrm{cut}}}
\newcommand{\ytt}{\ensuremath{y_{23}}}
\newcommand{\yttJ}{\ensuremath{y_{23}^\mathrm{J}}}
\newcommand{\yttD}{\ensuremath{y_{23}^\mathrm{D}}}
\newcommand{\kt}{\ensuremath{k_\mathrm{t}}}
\newcommand{\pt}{\ensuremath{p_\mathrm{t}}}
\newcommand{\ptone}{\ensuremath{p_{\mathrm{t}1}}}%
\newcommand{\pttwo}{\ensuremath{p_{\mathrm{t}2}}}%
\newcommand{\mt}{\ensuremath{m_\mathrm{t}}}

\newcommand{\JADE}{{\scshape jade}}

\newcommand{\Lthree}{{\scshape l}{\small 3}}

\newcommand{\LHC}{{\scshape lhc}}
\newcommand{\ATLAS}{{\scshape atlas}}
\newcommand{\CMS}{{\scshape cms}}
\newcommand{\ALICE}{{\scshape alice}}
\newcommand{\LHCb}{{\scshape lhc}{\scriptsize b}}

\newcommand{\Qslong}{\ensuremath{Q^2_\mathrm{L}}}
\newcommand{\Qsside}{\ensuremath{Q^2_\mathrm{side}}}
\newcommand{\Qsout}{\ensuremath{Q^2_\mathrm{out}}}

\newcommand{\Rlong}{\ensuremath{R_\mathrm{L}}}
\newcommand{\Rside}{\ensuremath{R_\mathrm{side}}}
\newcommand{\Rout}{\ensuremath{R_\mathrm{out}}}

\newcommand{\Rslong}{\ensuremath{R^2_\mathrm{L}}}
\newcommand{\Rsside}{\ensuremath{R^2_\mathrm{side}}}
\newcommand{\Rsout}{\ensuremath{R^2_\mathrm{out}}}

\newcommand{\yE}{\ensuremath{y_{\mathrm{E}}}}
\newcommand{\Ra}{\ensuremath{R_\mathrm{a}}}

\graphicspath{{.}{FIGS/}}                                  

\title{{The $\tau$-model of Bose-Einstein Correlations: \\ Some recent results}
        \footnote{Talk given at XLV International Symposium on Multiparticle Dynamics, Wildbad Kreuth, 4--9 Oct. 2015}}
\author{Wesley J. Metzger
\\
{IMAPP, Radboud University, 6525 AJ\ \ Nijmegen, The Netherlands}
\\
{W.Metzger@science.ru.nl}
}
 
\begin{document}
\maketitle
\abstract{%
   Bose-Einstein correlations of pairs of identical charged pions
   produced in hadronic Z decays and in 7\,\TeV\ pp minimum bias interactions are
   investigated within the framework of the \taumodel.
}
\section{Introduction} \label{intro}
After a brief review of relevant previous results,
new \textit{preliminary\/} results are presented on the dependence of
the Bose-Einstein correlation function on track and jet multiplicity, pair transverse momentum and pair rapidity,
using a parametrization
which has been found~\cite{L3_levy:2011} to describe well
Bose-Einstein correlations in hadronic \PZ\ decay, namely that of
the \taumodel~\cite{Tamas;Zimanji:1990,ourTauModel}.
 
\subsection{`Classic' Parametrizations}    \label{review}
The    Bose-Einstein correlation function, $R_2$, is often parametrized as
\begin{equation} \label{eq:gauss_param}
  R_2 =    \gamma \left[ 1+ \lambda \exp \left(-\left(rQ\right)^{2} \right) \right] (1+ \epsilon Q) \;,
\end{equation}
and is measured by $R_2(Q)=\rho(Q)/\rho_0(Q)$,
where $\rho(Q)$ is the density of identical boson pairs with invariant four-momentum difference
$Q=\sqrt{-(p_1-p_2)^2}$
and $\rho_0(Q)$ is the similar density in an artificially constructed reference sample,
which should differ from the data only in that it does not contain the effects of Bose symmetrization of identical bosons.

\subsection{\taumodel}     \label{taumodel}
However, the ``classic'' parametrization of \Eq{eq:gauss_param} is found to be inadequate, even when it is
generalized to allow for a L\'evy distribution of the source:
\begin{equation} \label{eq:slevy_param}
  R_2 =    \gamma \left[ 1+ \lambda \exp \left(-\left(rQ\right)^{\alpha} \right) \right] (1+ \epsilon Q) \;, \quad 0<\alpha\le2
\end{equation}
This was not realized for a long time because the correlation function was only plotted up to $Q=2\,\GeV$
or less.  In Ref.~\citen{L3_levy:2011} $Q$ was plotted to 4\,\GeV, and it became apparent that there is a
region of anti-correlation ($R_2<1$) extending from about $Q=0.5$ to $1.5\,\GeV$.  This anti-correlation,
which one might term Bose-Einstein Anti-Correlations (BEAC),
as well as the Bose-Einstein correlations (BEC) are well described by the \taumodel.

In the \taumodel\ $R_2$ is found to depend not only on $Q$, but also
on quantities $a_1$ and $a_2$.
For two-jet events $a=1/\mt$, where
$\mt=\sqrt{m^2+\pt^2}$ is the transverse mass of a particle).
Parameters of the model are the parameters of the L\'evy distribution which describes the proper time of particle emission:
$\alpha$, the index of stability of the L\'evy distribution; a width parameter $\Delta\tau$; and the proper time $\tau_0$ at which
particle production begins.
 
We shall use a simplified parametrization~\cite{L3_levy:2011} where $\tau_0$ is assumed to be zero and
 $a_1$ and $a_2$ are combined with $\Delta\tau$ to form an effective radius $R$:
\begin{subequations} \label{eq:asymlevR2}
\begin{align}
  \begin{split}
    R_2(Q) &= \gamma \left[ 1+ \lambda \cos \left(\left(R_\mathrm{a}Q\right)^{2\alpha} \right)
             \exp \left(-\left(RQ\right)^{2\alpha} \right) \right] \\
           & \quad \cdot  (1+ \epsilon Q) \;,
           \label{eq:asymlevR2_}   \raisetag{\bigskipamount}
  \end{split}
\\
    R_\mathrm{a}^{2\alpha} &= \tan\left(\frac{\alpha\pi}{2}\right) R^{2\alpha} \;.     \label{eq:asymlevRaR}
\end{align}
\end{subequations}
Note that the difference between the parametrizations of \Eqs{eq:slevy_param} and \ref{eq:asymlevR2}
is the presence of the cosine term, which provides the description of the anti-correlation.
The parameter $R$ describes the BEC peak, and $R_\mathrm{a}$ describes the anti-correlation region.
While one might have had the insight to add, \adhoc, a $\cos$ term to \Eq{eq:slevy_param}, it is the \taumodel\
which provides a physical reason for it and
which predicts a relationship,     \Eq{eq:asymlevRaR}, between $R$ and $R_\mathrm{a}$.
 
\begin{figure}[ht] \centering
  \includegraphics*[width=.35\textwidth,clip,viewport=56 87  518 680]{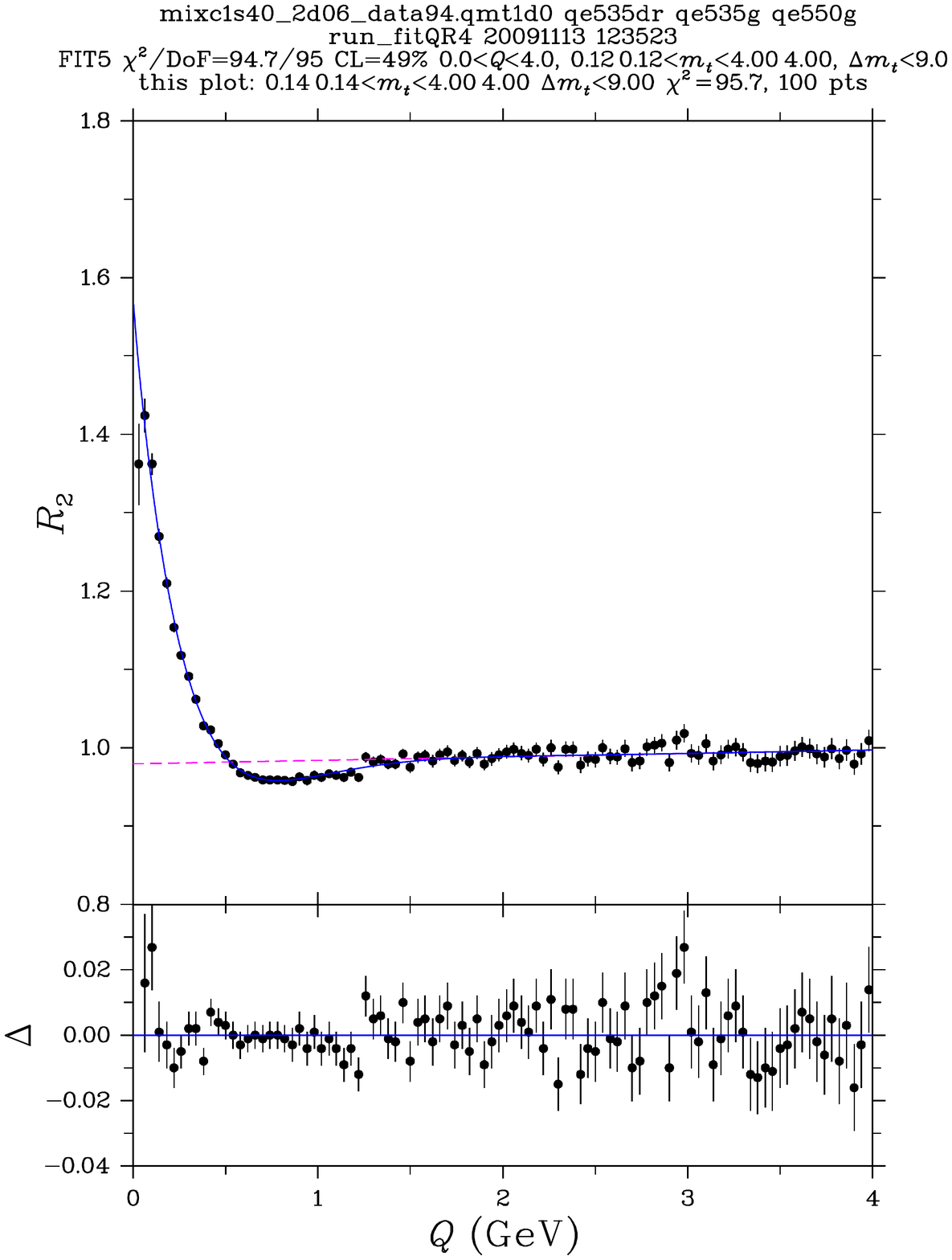}
  \caption{The Bose-Einstein correlation function $R_2$ for two-jet events.
           The curve corresponds to the fit of \Eq{eq:asymlevR2}.
           Also plotted is $\Delta$, the difference between the fit and the data.
           The dashed line represents the long-range part of the fit, \ie, $\gamma(1+\epsilon Q)$.
           The figure is taken from Ref.~\citen{L3_levy:2011}.
          }
  \label{fig:L32jet}
\end{figure}
A fit of \Eq{eq:asymlevR2} to \Lthree\ two-jet events is shown in \Fig{fig:L32jet}, from which
it is seen that the \taumodel\ describes both the BEC peak and the BEAC dip quite well.
Also the three-jet data are well described~\cite{L3_levy:2011},
which is perhaps surprizing
since the \taumodel\ is inspired by a picture of fragmentation of a single string.

It must also be pointed out that the \taumodel\ has its shortcomings:
The \taumodel\ predicts that $R_2$ depends on the two-particle
momentum difference only through $Q$, not through components of $Q$.
However, this is found not to be the case~\cite{L3_levy:2011}.
Nevertheless, regardless of the validity of the \taumodel, \Eq{eq:asymlevR2} provides a good description of the data.
Accordingly, we shall use it in most of the following.

\subsection{Data}  \label{data}
The \Pep\Pem\ data were collected by the
\Lthree\ detector 
at a center-of-mass energy of $\sqrt{s}\simeq 91.2$\,\GeV\kern-0.1em.
Approximately 36 million like-sign pairs of well-measured charged tracks from about 0.8 million
hadronic Z decays are used. 
This data sample is identical to that of Ref.~\citen{L3_levy:2011}.
The same event mixing technique is used to construct $\rho_0$ as in  Ref.~\citen{L3_levy:2011}.
 
The minimum-bias pp data were collected by the \ATLAS\ detector
at a center-of-mass energy of $\sqrt{s}=7$\,\TeV\kern-0.1em.
The sample contains approximately $1.8\times10^9$ like-sign charged track pairs from approximately $10^7$ events.
The acceptance in pseudorapidity is $\abs{\eta}<2.5$.
This data sample is identical to the 7\,\TeV\kern-0.1em\ sample used in the recent \ATLAS\ paper on BEC~\cite{ATLAS:be1}.
However the new results reported here, which are taken from a Ph.D.\ thesis~\cite{Astalos:thesis},
use, unless otherwise stated, a reference sample constructed by the opposite-hemisphere (OHP) method where the
three-momentum, $\vec{p}$ of one of the particles of the pair is replaced by $-\vec{p}$.
The recent \ATLAS\ paper~\cite{ATLAS:be1} used unlike-sign pairs to construct $\rho_0$, which is unsuitable
for studying the region of anti-correlation, since in that region the $\rho^0$ meson is a dominant feature.

The Durham and \JADE\ algorithms are used to classify \Pee\ events according to the number of jets.
The number of jets in a particular event depends on the jet resolution parameter of the algorithm, \ycut.
We define \yttD\ (\yttJ) as the value of \ycut\ at which the number of jets in the event changes from two to three using the
respective algorithms.
Small \ytt\ corresponds to narrow two-jet events, large \ytt\ to events with three (or more) well-separated jets.
The two algorithms lead to similar results.
 
\section{New \textit{Preliminary}\/ Results}                                  \label{NewResults}
\begin{figure*}     \centering
\begin{minipage}{.32\textwidth}
  \centering
  \includegraphics[width=1.00\textwidth,clip,viewport=47 9 536 412]{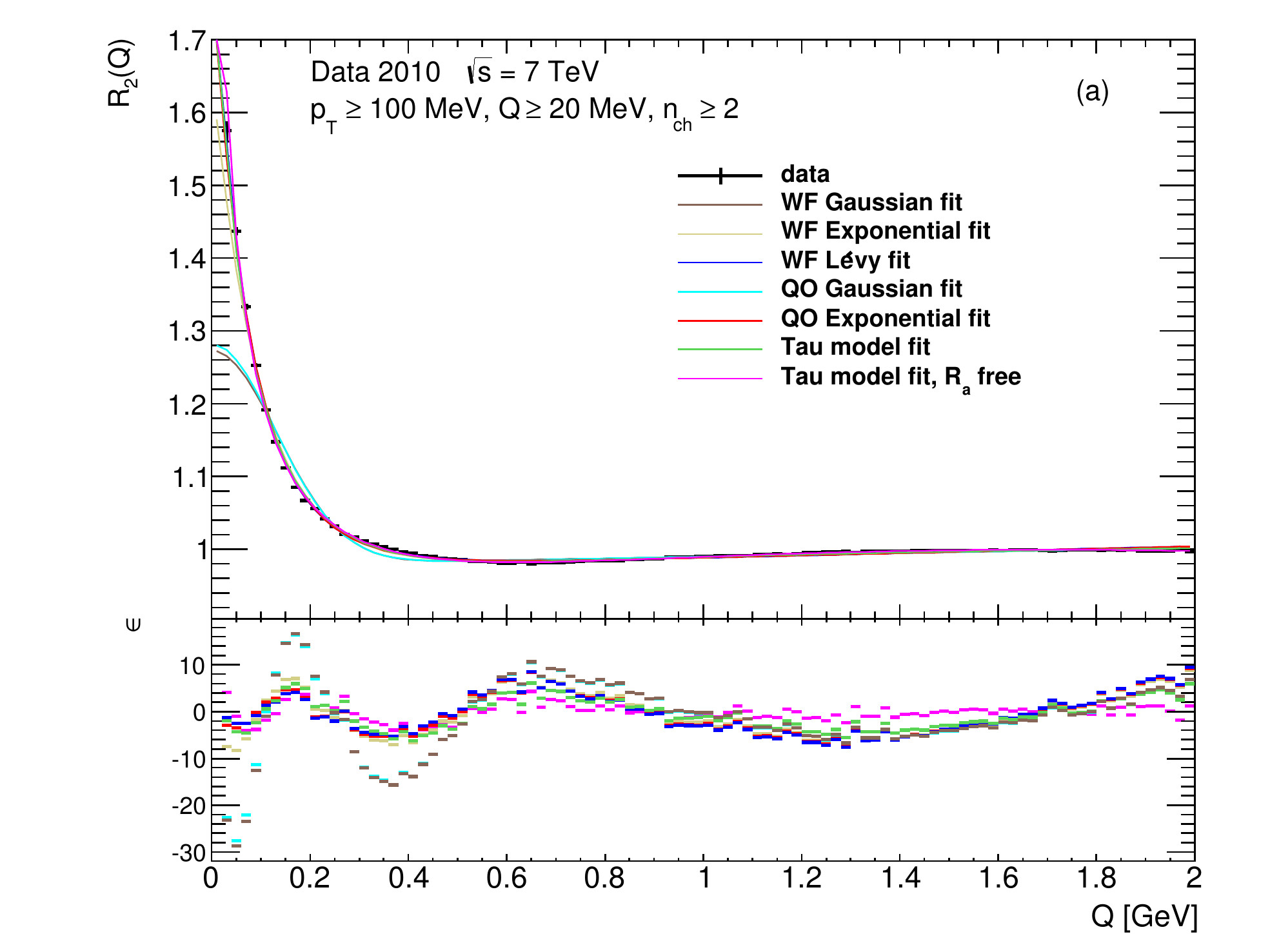}
\end{minipage}
  \begin{minipage}{.32\textwidth}
    \includegraphics[width=1.00\textwidth,clip,viewport=47 9 539 412]{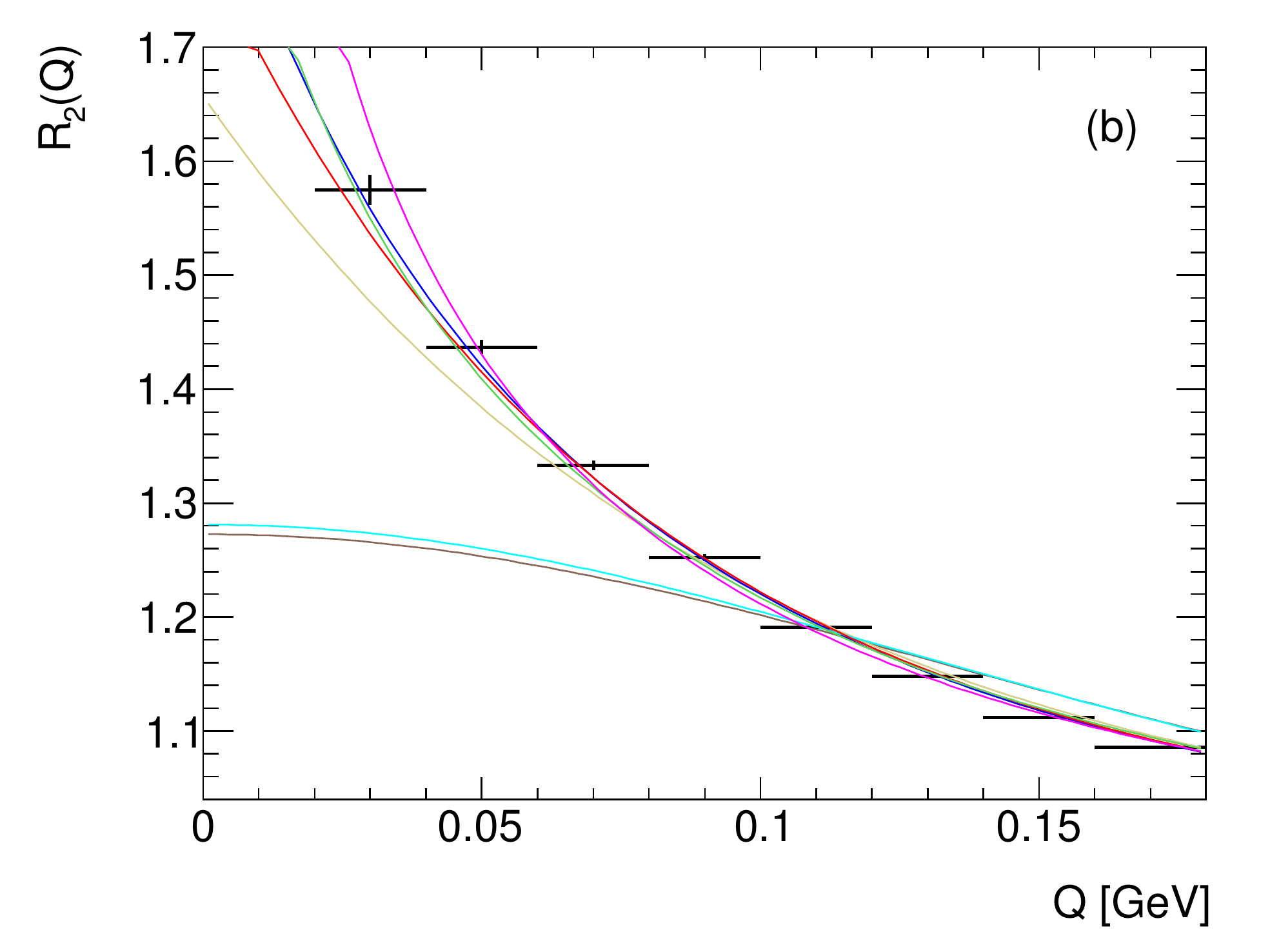}
  \end{minipage}
  \begin{minipage}{.32\textwidth}
    \includegraphics[width=1.00\textwidth,clip,viewport=47 9 539 412]{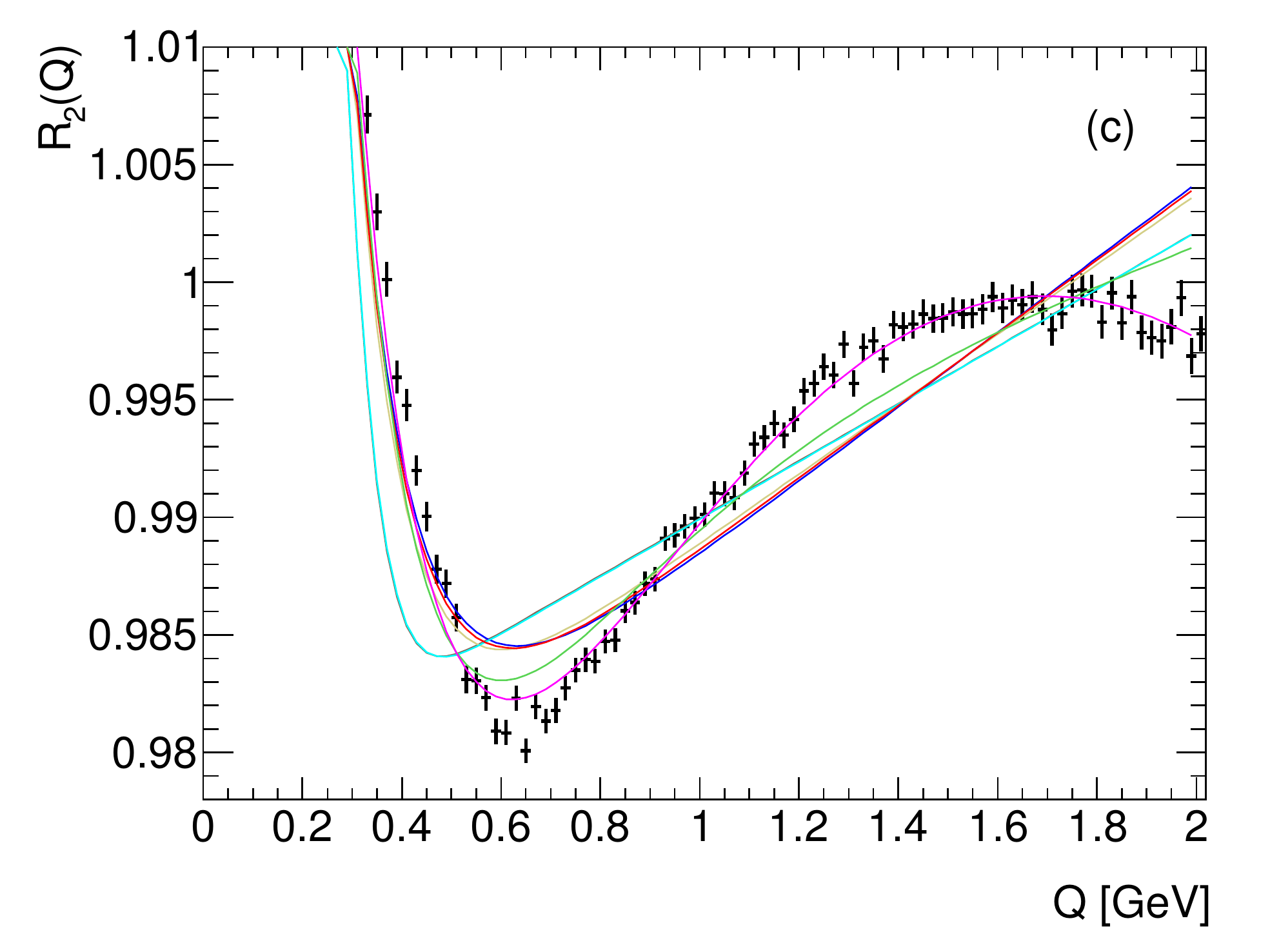}
  \end{minipage}
  \caption{$R_2$ for 7\,\TeV\ pp minimum bias interactions with the results of fits of several
           parametrizations~\cite{Astalos:thesis}.}
  \label{fig:ppR2}
\end{figure*}

As in \Pee\ annihilation \cite{L3_levy:2011} , the pp minimum bias data are found \cite{Astalos:thesis} to be best described by
\Eq{eq:asymlevR2_}. While other parametrizations can describe the BEC peak,
only this parametrization (with $R_\text{a}$ a free parameter) comes close to describing not only the BEC peak but also the BEAC
region, as is seen in Fig.~\ref{fig:ppR2}.  Previously, \CMS\ had also observed the BEAC and fit it with the
\taumodel\ parametrization~\cite{CMS:be2}.
 
%
 
\begin{figure}     \centering
    \includegraphics[width=.43\textwidth,clip,viewport=0 0 550 412]{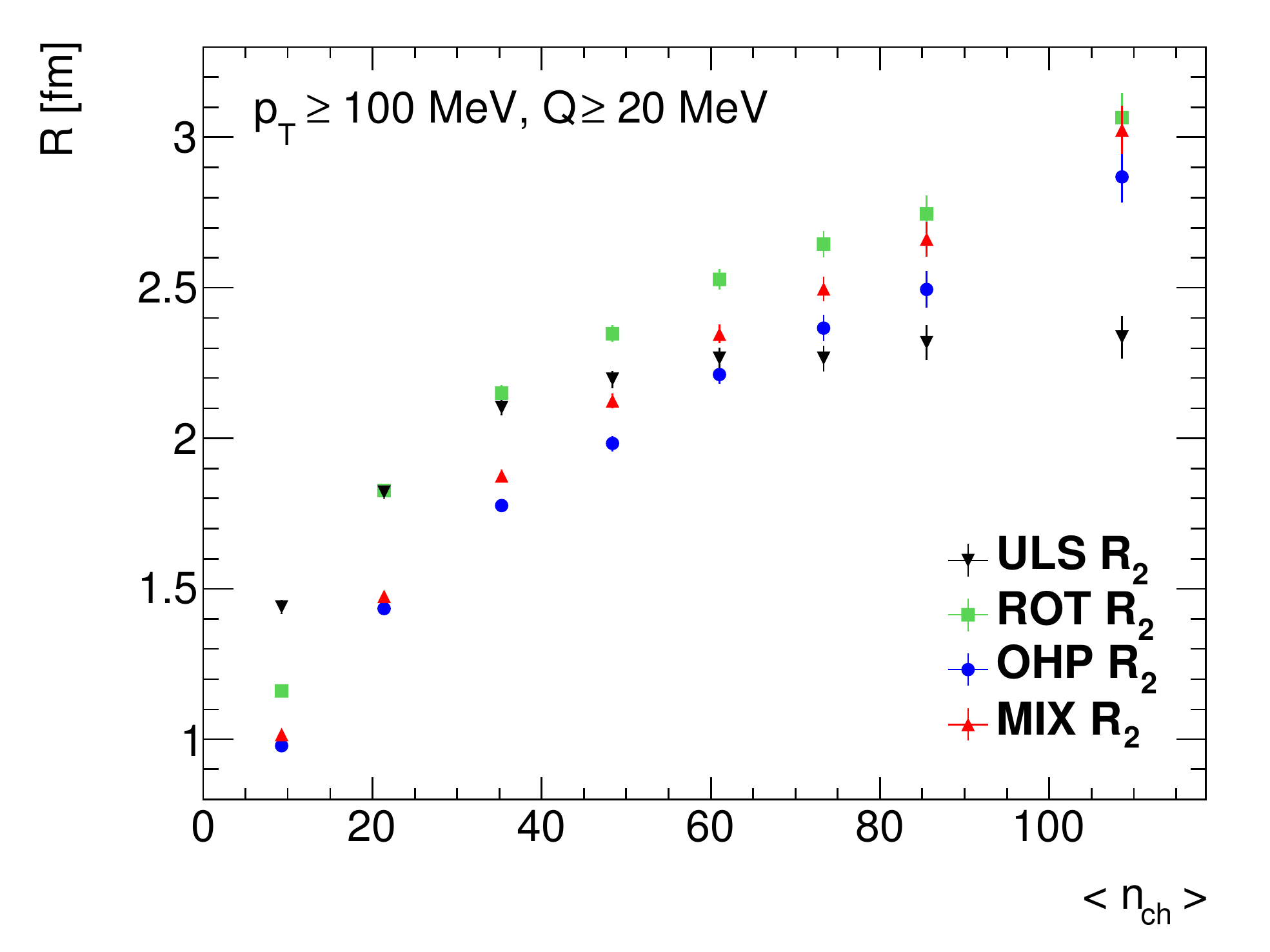}
 
    \includegraphics[width=.43\textwidth,clip,viewport=0 0 550 412]{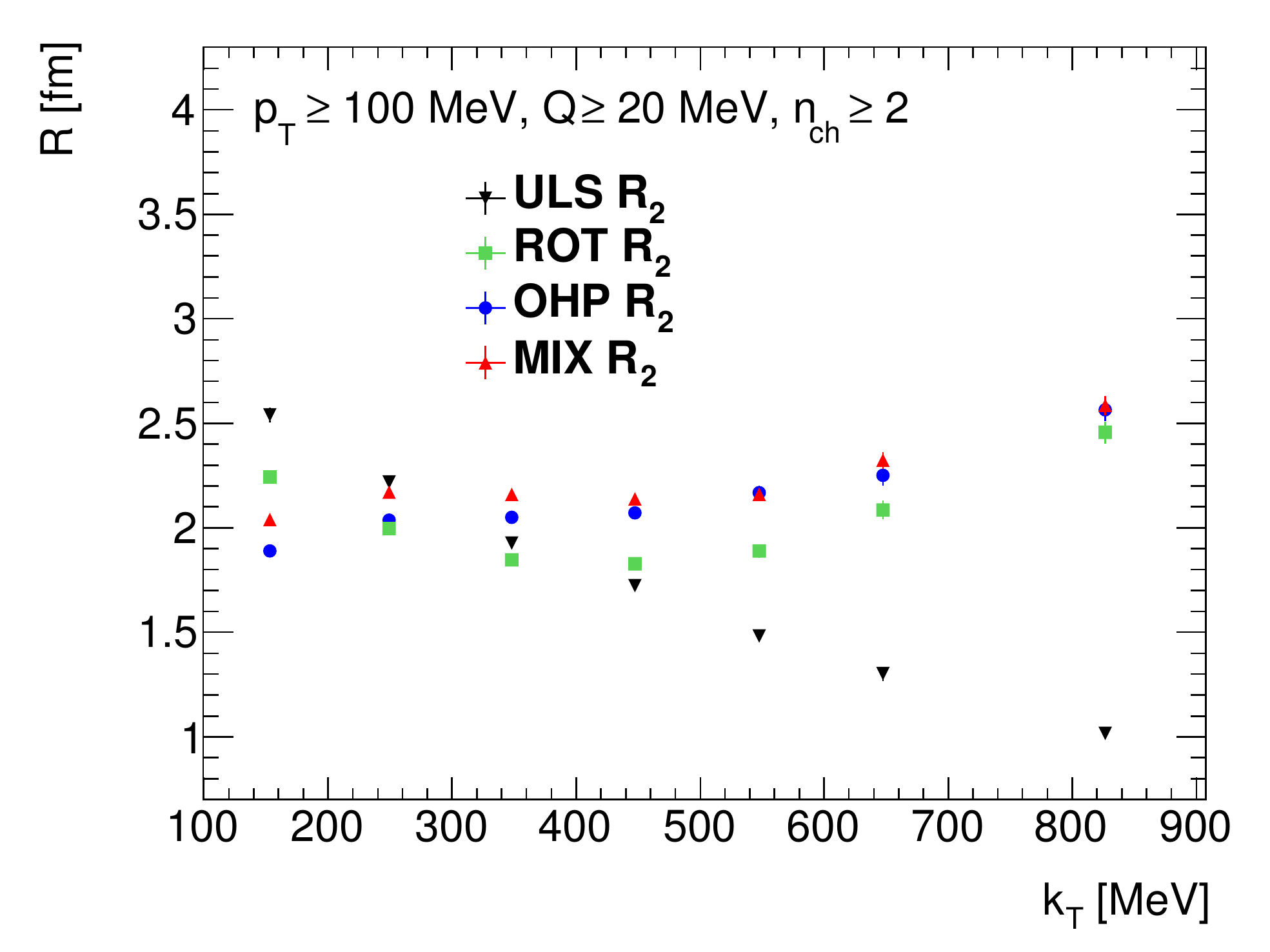}
  \caption{The dependence of $R$ (top) on $N$  and (bottom) on \kt\ with respect to the beam axis for different reference samples
           in 7\,\TeV\ pp minimum bias events.  The exponential parametrization is used~\cite{Astalos:thesis}.
           \label{fig:refsamp}
          }
\end{figure}

As has frequently been pointed out (see, \eg, Ref.~\citen{wes:ismd2014}) the values of the parameters of BEC depend strongly on the
choice of reference sample.  This is seen again in Fig.\,\ref{fig:refsamp}, where the dependence of $R$ on the charged
track multiplicity, $N$, and on the average pair transverse momentum, $\kt=(\vec{\ptone}+\vec{\pttwo})/2$, is shown for the
exponential parametrization of $R_2$, \ie, \Eq{eq:slevy_param} with $\alpha=1$, which is the parametrization used in the
\ATLAS\ paper \cite{ATLAS:be1}.  The dependences with unlike-sign (ULS) reference sample are markedly different from those with the
other samples.  Whereas $R$ increases somewhat less than linearly with $N$ for the other reference samples, it appears to saturate for the
ULS sample.  And, whereas $R$ decreases approximately linearly with \kt\ for the ULS sample, for the rotated reference sample it
decreases only up to $\kt\approx450\,\MeV$ and then increases, and for the other reference samples it is first roughly constant with
\kt\ and then increases.
 
An unfortunate property of the \taumodel\ parameterization  is that
the estimates of $\alpha$, $R$, and \Ra\ from the fits tend to be highly correlated.
Therefore, when studying the dependence of $R$ on quantities such as $N$ or \kt, in order to stabilize the fits $\alpha$ is fixed to
the value obtained in a fit to all events.
In examining the dependence of $R$ on charged track multiplicity  
and on \kt, the parametrization of
\Eq{eq:asymlevR2_} is used (with \Ra\ given by \Eq{eq:asymlevRaR} for \Pee\ and a free parameter for pp).
 
\begin{figure}     \centering
        \includegraphics[width=0.34\textwidth,angle=270,clip,viewport=114 114 556 717]{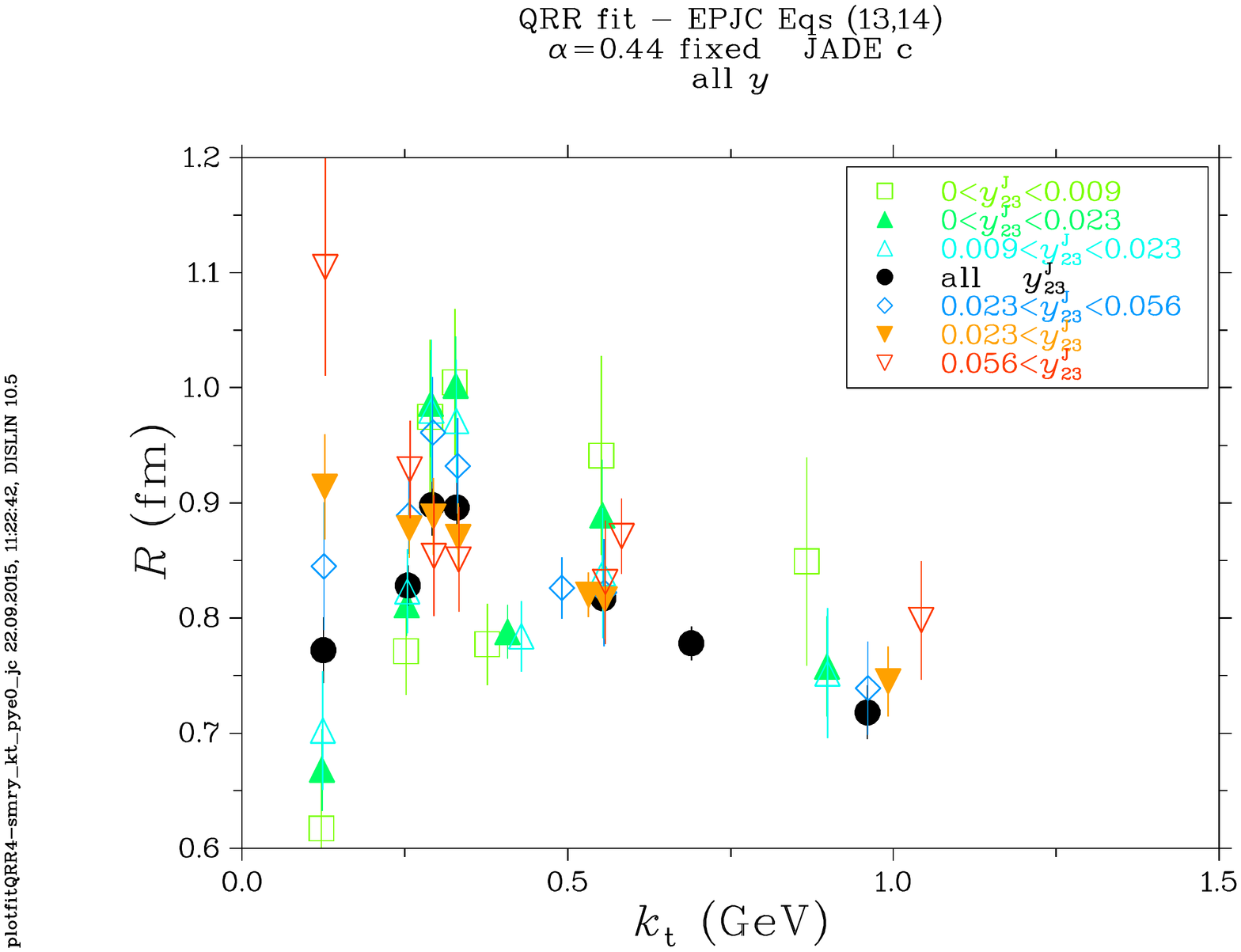}
    \includegraphics[width=0.48\textwidth,clip,viewport=0 0 540 412]{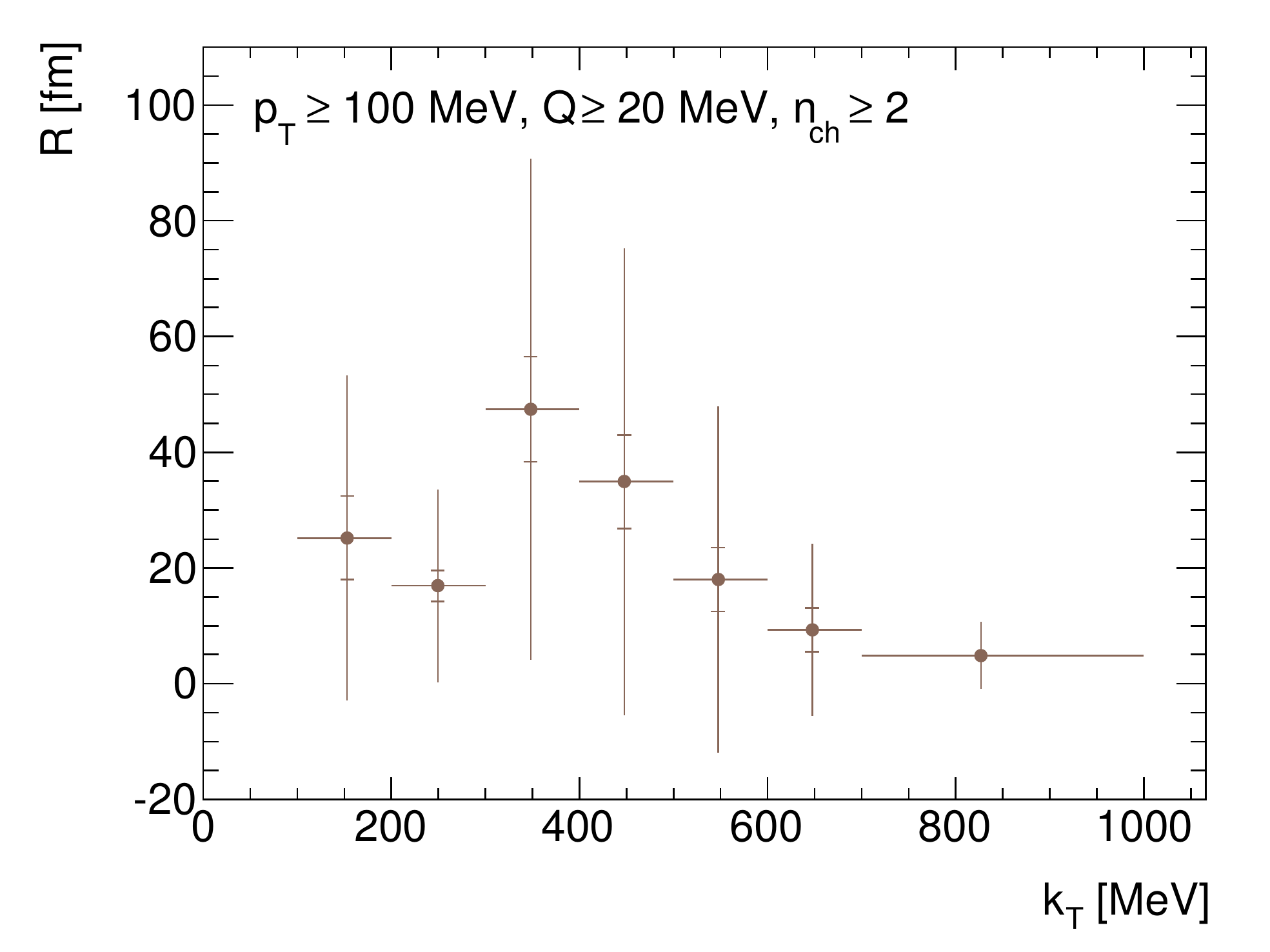}
  \caption{The dependence of $R$ on \kt\ for \Pee\ (top) and pp (bottom).
           The \Pee\ results are shown for various selections on \yttJ.
           The \taumodel\ parametrization is used with \Ra\ a free parameter.
           \label{fig:kt}
          }
\end{figure}

The dependence of $R$ on \kt\ with respect to the thrust axis
for \Pee\ and with respect to the beam axis for pp is shown in \Fig{fig:kt} for the \taumodel\ parametrization with \Ra\ a free
parameter is shown in \Fig{fig:kt}.  The behavior of two-jet \Pee\ seems similar to pp, although the error bars are large.
 
\begin{figure} \centering
   \includegraphics[width=0.33\textwidth,angle=270,clip,viewport=124 122 554 717]{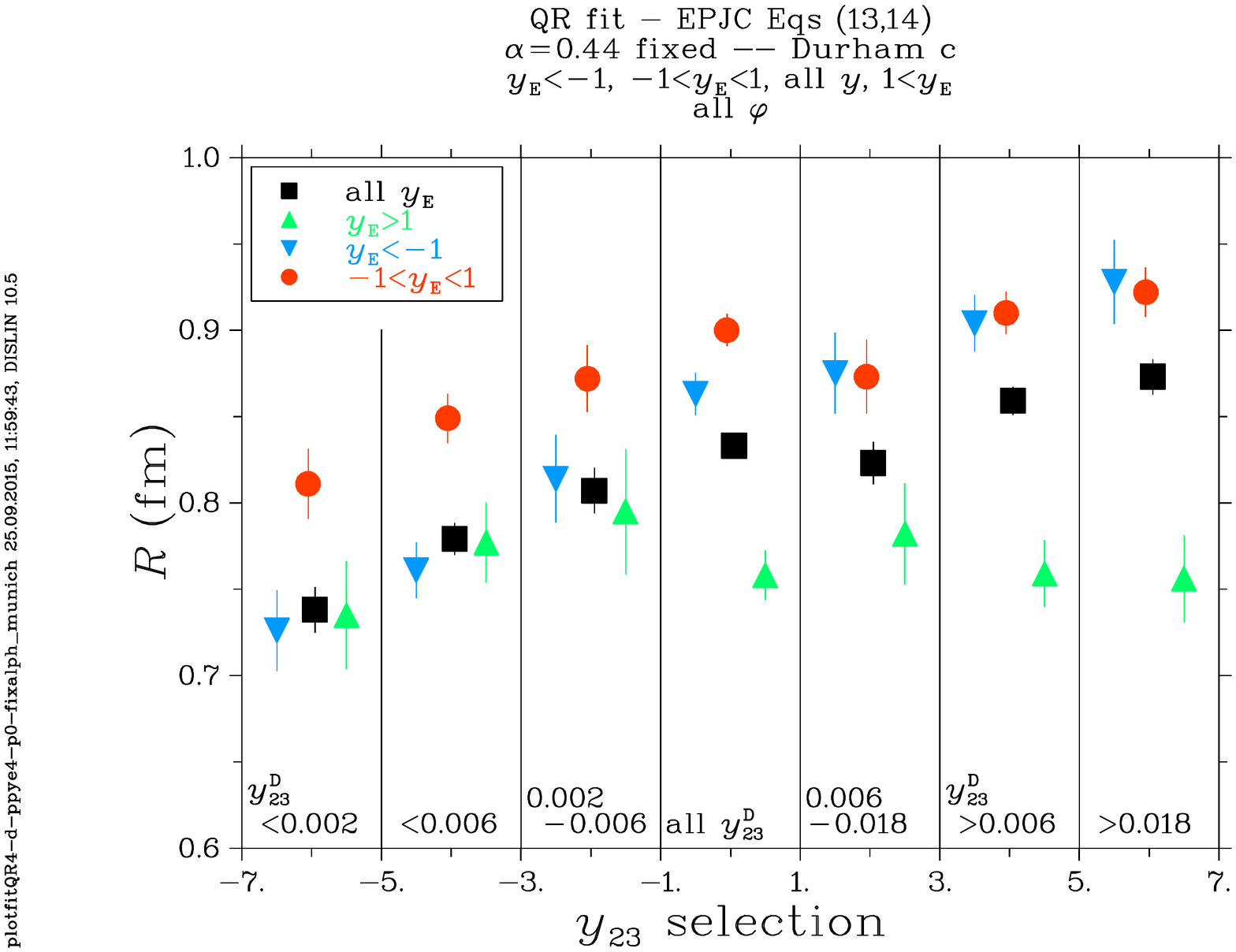}
  \caption{$R$ obtained in          fits of \Eq{eq:asymlevR2}
           for various rapidity and \yttD\ intervals.
           \label{fig:rap}
          }
\end{figure}

For both \Pee\ and pp, $R$ is found \cite{wes:ismd2014,Astalos:thesis} to increase with multiplicity (not shown).
Further, for \Pee\ $R$ is found \cite{wes:ismd2014} to increase with `jettiness', as measured by \ytt.
 
In \Pee, $R$ is also found to depend on the rapidity of the pair, as shown in \Fig{fig:rap}.
Here the rapidity, \yE, is defined with respect to the thrust axis with the positive thrust axis chosen to be in the same
hemisphere as the most energetic jet.  Then $\yE>1$ selects almost always tracks from the most energetic quark jet, and $\yE<-1$
selects mostly tracks from the other quark jet with the contribution of tracks from the gluon jet increasing as \ytt\ increases,
\ie, as the events become more three-jetlike.
The intermediate \yE\ region contains tracks from the gluon jet and low-energy
tracks from both quark jets.  One observes that $R$ is roughly independent of \ytt\ for $\yE>1$. This value of $R$ is also found for
$\yE<-1$  in the case of two-jet events.  These are the situations of `pure' quark jets.  As \ytt\ increases $R$ for $\yE<-1$ also
increases, reflecting the increasing contribution of tracks from the gluon.  The region $-1<\yE<1$ has a larger value of $R$ for
two-jet events, and this value increases with \ytt. For three-jet events, where the gluon contributes to both $-1<\yE<1$ and
$\yE<-1$, the values of $R$ are approximately equal.
 
\begin{figure} \centering
\begin{minipage}{.49\textwidth}
  \centering
  \includegraphics[width=.65\textwidth,angle=270,clip,viewport=121 121 553 716]{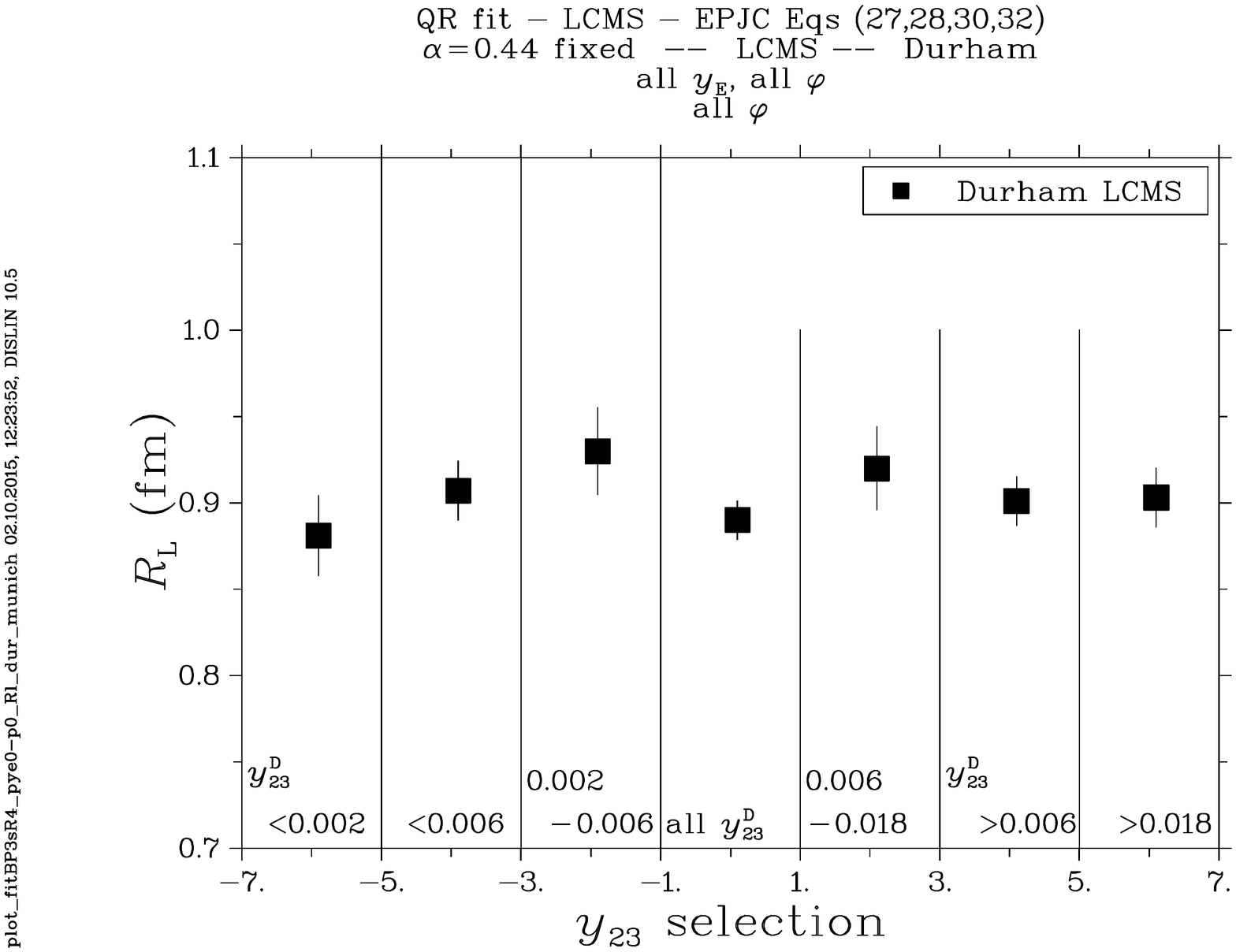}
\end{minipage}
\begin{minipage}{.49\textwidth}
  \centering
  \includegraphics[width=.65\textwidth,angle=270,clip,viewport=121 121 553 716]{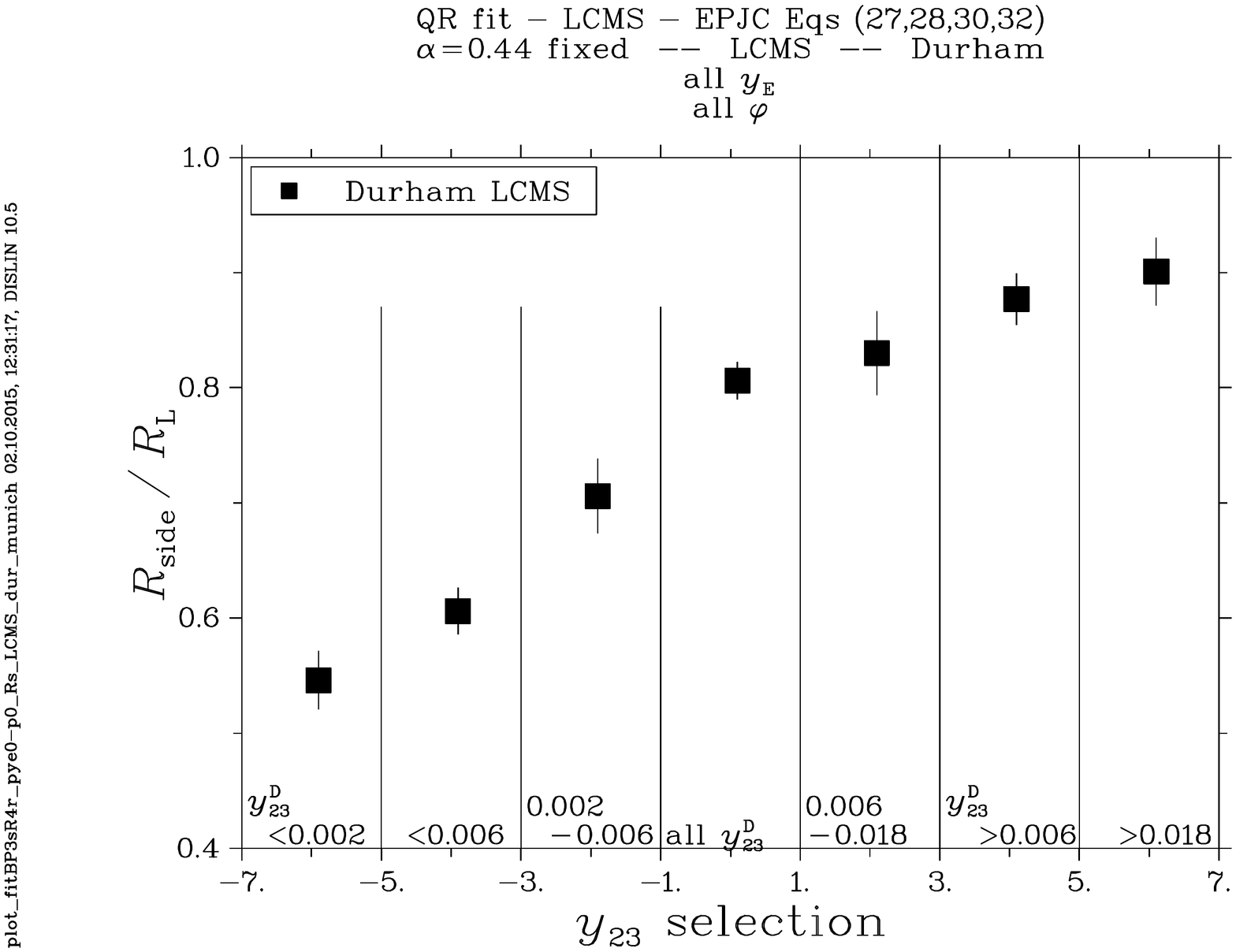}
\end{minipage}
  \caption{\Rlong\ and $\Rside/\Rlong$ obtained in          fits of \Eq{eq:asymlevR2}
           for various rapidity and \yttD\ intervals.
           \label{fig:long-side}
          }
\end{figure}

As mentioned above, the \taumodel\ is known to break down.  In the \taumodel\ $R_2$ depends on $Q$, and not on components
of $Q$.  This was tested in Ref.~\citen{L3_levy:2011}.  In \Eq{eq:asymlevR2} $R^2Q^2$ is replaced by
  $\Rslong\Qslong + \Rsside\Qsside + \Rsout\Qsout$ and fits for two-jet events performed in the LCMS.  Different values were found
for \Rlong, \Rout, and \Rside.  Here results are shown in \Fig{fig:long-side} for \Rlong\ and \Rside\
for various selections of \yttD.  One sees that \Rlong\ is independent of \ytt, while \Rside\ increases with `jettiness'.
Thus the increase in $R$ occurs mainly in the transverse plane.
This is consistent with the increase in $R$ being due to the increasing hardness of the gluon, which was seen in the rapidity
dependence above.

\begin{figure*}              \centering
\begin{minipage}{.32\textwidth}
    \includegraphics*[width=0.99\textwidth,clip,viewport=47 9 536 412]{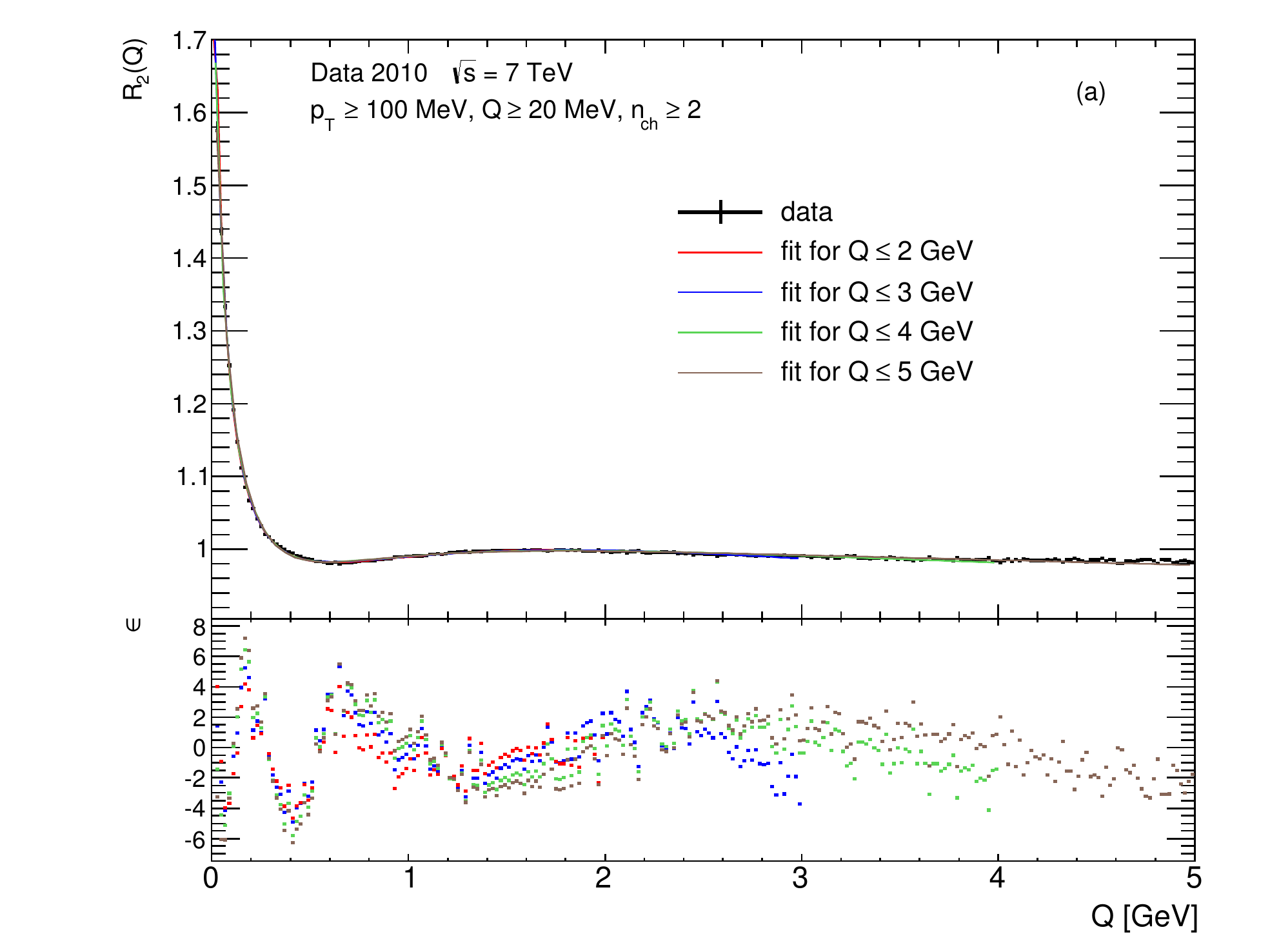}
\end{minipage}
\hfill
\begin{minipage}{.32\textwidth}
    \includegraphics*[width=1.00\textwidth,clip,viewport=45 9 540 411]{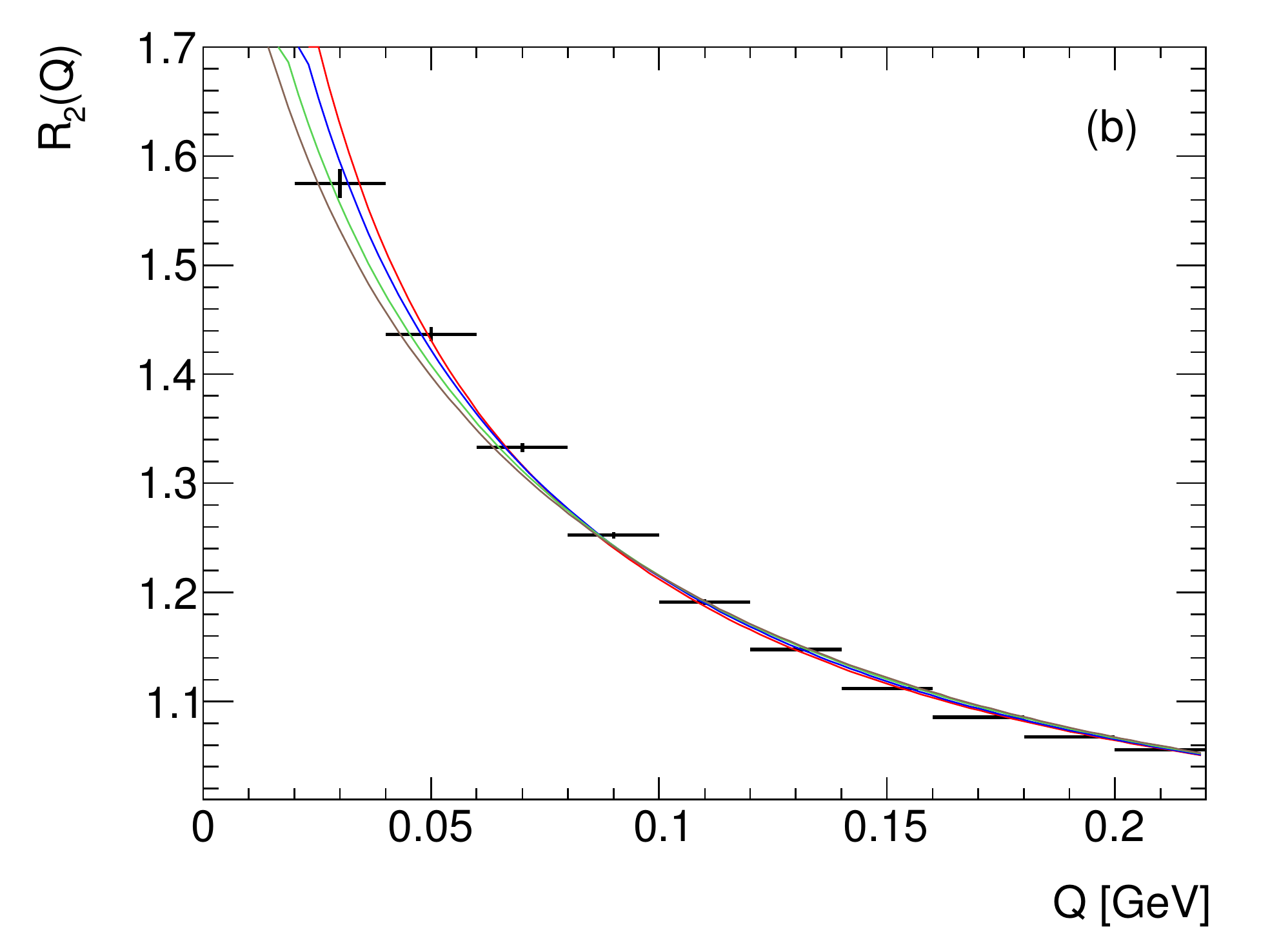}
\end{minipage}
\hfill
\begin{minipage}{.32\textwidth}
    \includegraphics*[width=1.00\textwidth,clip,viewport=45 9 540 411]{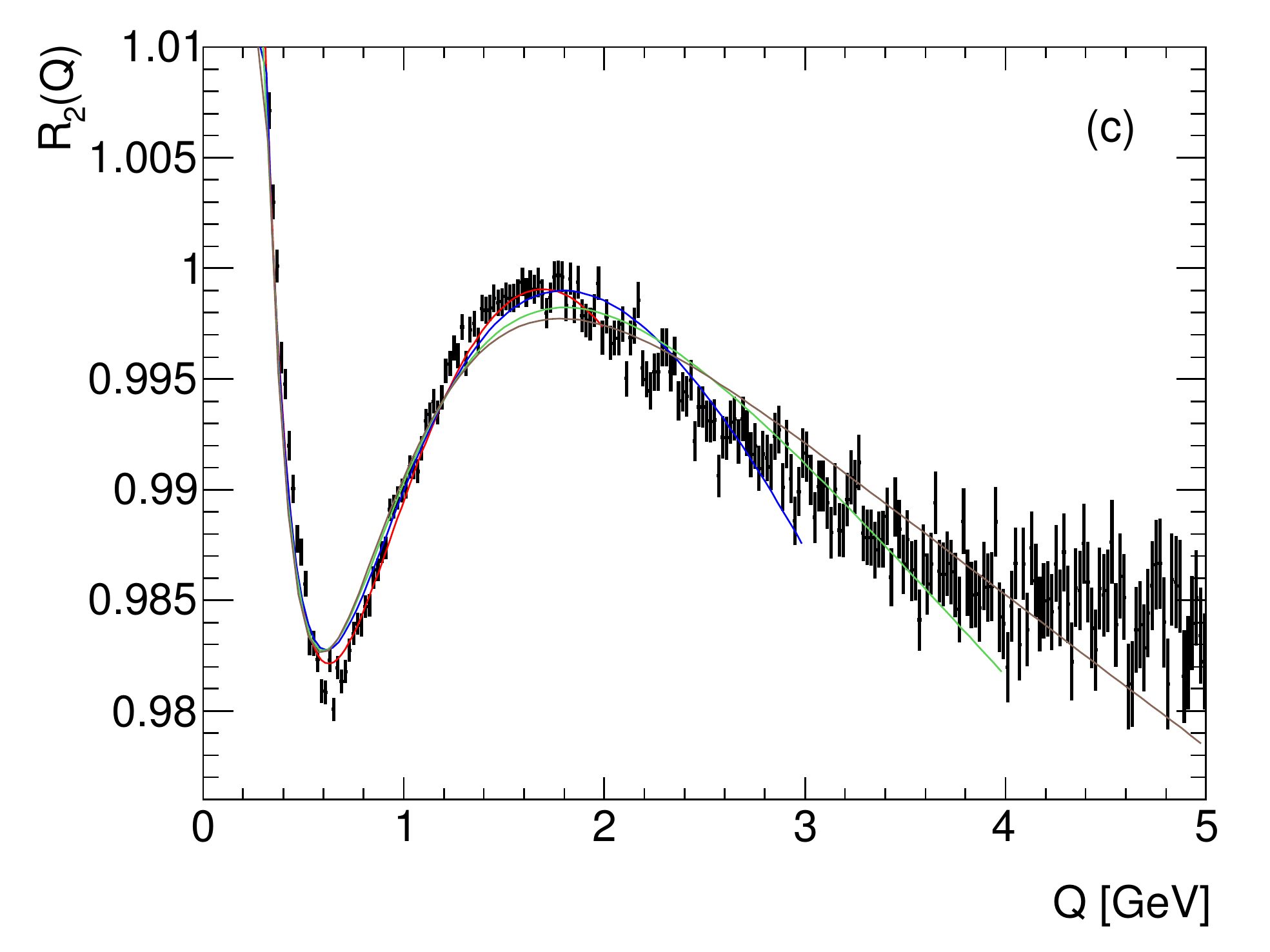}
\end{minipage}
  \caption{Fits of \Eq{eq:asymlevR2_}, with \Ra\ a free parameter, to 7\,\TeV\ minimum bias data
           for various choices of upper limit of the fit range~\cite{Astalos:thesis}.
          }
  \label{fig:QU}
\end{figure*}

A large source of systematic uncertainty on the parameter values
is the range of $Q$ over which the fit is performed.
A higher upper limit, $Q_\text{U}$, of the fit range gives more data to determine the baseline,
which in turn affects the values of the parameters describing BEC and BEAC.
Clearly, $Q_\text{U}$ should be chosen well beyond the anti-correlation region.
\Fig{fig:QU} and \Tab{tab:QU} show this effect.
The values of the parameters change drastically when the upper limit, $Q_\text{U}$, is increased from 2 to 3\,\GeV.
As $Q_\text{U}$ is further increased the change is less, but still large.  It is clear that the parametrization does not
fully describe the data.  Nevertheless, it is much better than all the other parametrizations that were tried.
It is also worth noting that the value of $\alpha$ is quite different from the value $\alpha = 0.41\pm0.02^{+0.04}_{-0.06}$
found for two-jet \Pee\ events \cite{L3_levy:2011}.
 
\begin{table*}
  \caption{Results of fits of \Eq{eq:asymlevR2_}, with \Ra\ a free parameter, for various choices of upper limit, $Q_\text{U}$, of
           the fit range~\cite{Astalos:thesis}.}
  \label{tab:QU}
  \centering
\definecolor{brown}{rgb}{.645,.164,.164}
  {
   \begin{tabular}{|ccccc|}  \hline
      $Q_\mathrm{U}$ &\color{red} 2\,\GeV &\color{blue} 3\,\GeV &\color{green} 4\,\GeV  &\color{brown} 5\,\GeV       \\
     \hline
      $\alpha$             &$0.108\pm0.001$&$0.186\pm0.005$&$0.235\pm0.003$ &$0.261\pm0.003$ \\
      $R$ (fm)             &$17.8\pm0.7   $&$6.7\pm0.5$    &$4.1 \pm0.2   $ &$3.3 \pm0.1 $   \\
      \Ra\ (fm)            &$43.4\pm1.2 $  &$3.0 \pm0.2 $  &$1.80\pm0.04$   &$1.52\pm0.02$   \\
      $\lambda$            &$3.08\pm0.05$  &$1.91\pm0.10$  &$1.36\pm0.05$   &$1.15\pm0.03$   \\
      \hline
   \end{tabular}
  }
\end{table*}

\section{BEAC in more detail}  \label{sect:BEAC}
In the \taumodel\ BEAC arises through the correlation of coordinate space and momentum space.
Recently, another explanation has been proposed, namely the non-zero size of the pion \cite{Bialas:2013oza, Bialas:2015hfa}.
A detailed investigation of the BEAC region seems therefore warranted, in the hope of distinguishing between these explanations.

Since fits of \Eq{eq:asymlevR2_}, with \Ra\ a free parameter, provide a reasonable description of the anti-correlation region, they
are used to show how BEAC depend on `jettiness' in \Pee\ and to compare the dependence of BEAC on multiplicity and \kt\
in two-jet \Pee\ and pp minimum bias events.
 
\begin{figure} \centering
  \includegraphics[width=0.43\textwidth,clip,viewport=46 57 496 664]{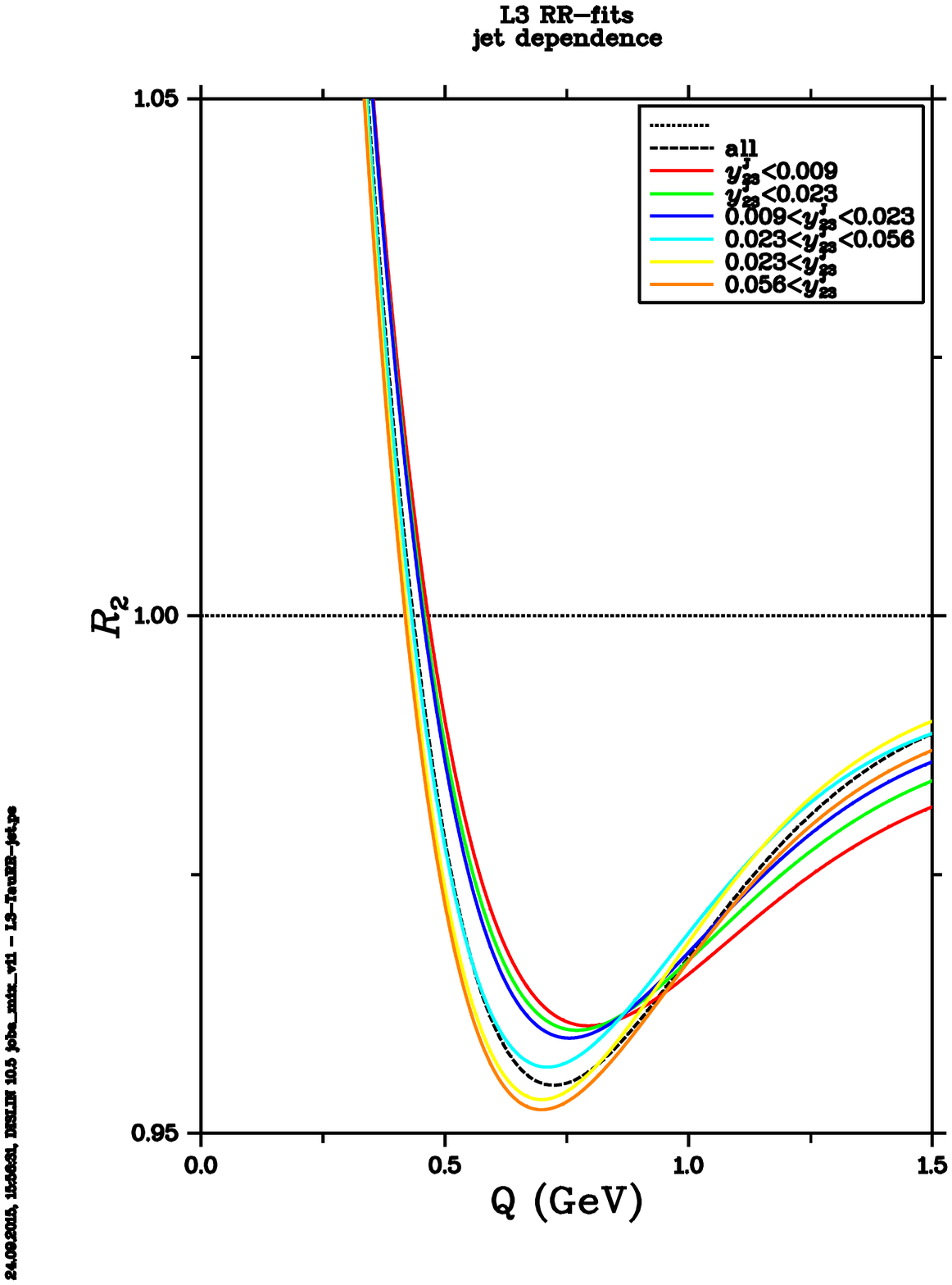}
  \caption{The result of fits of \Eq{eq:asymlevR2} to \Pee\ for various \yttJ\ selections.
           \label{fig:BEACjet}
          }
\end{figure}

\Fig{fig:BEACjet} shows the dependence of the BEAC on \yttJ.  It is seen that the anti-correlation dip becomes deeper and its
minimum moves slighlty lower in $Q$ as the `jettiness' increases, i.e., as one moves from two- to three-jet events.

\begin{figure*} \centering
  \includegraphics[width=0.43\textwidth,clip,viewport=47 59 496 653]{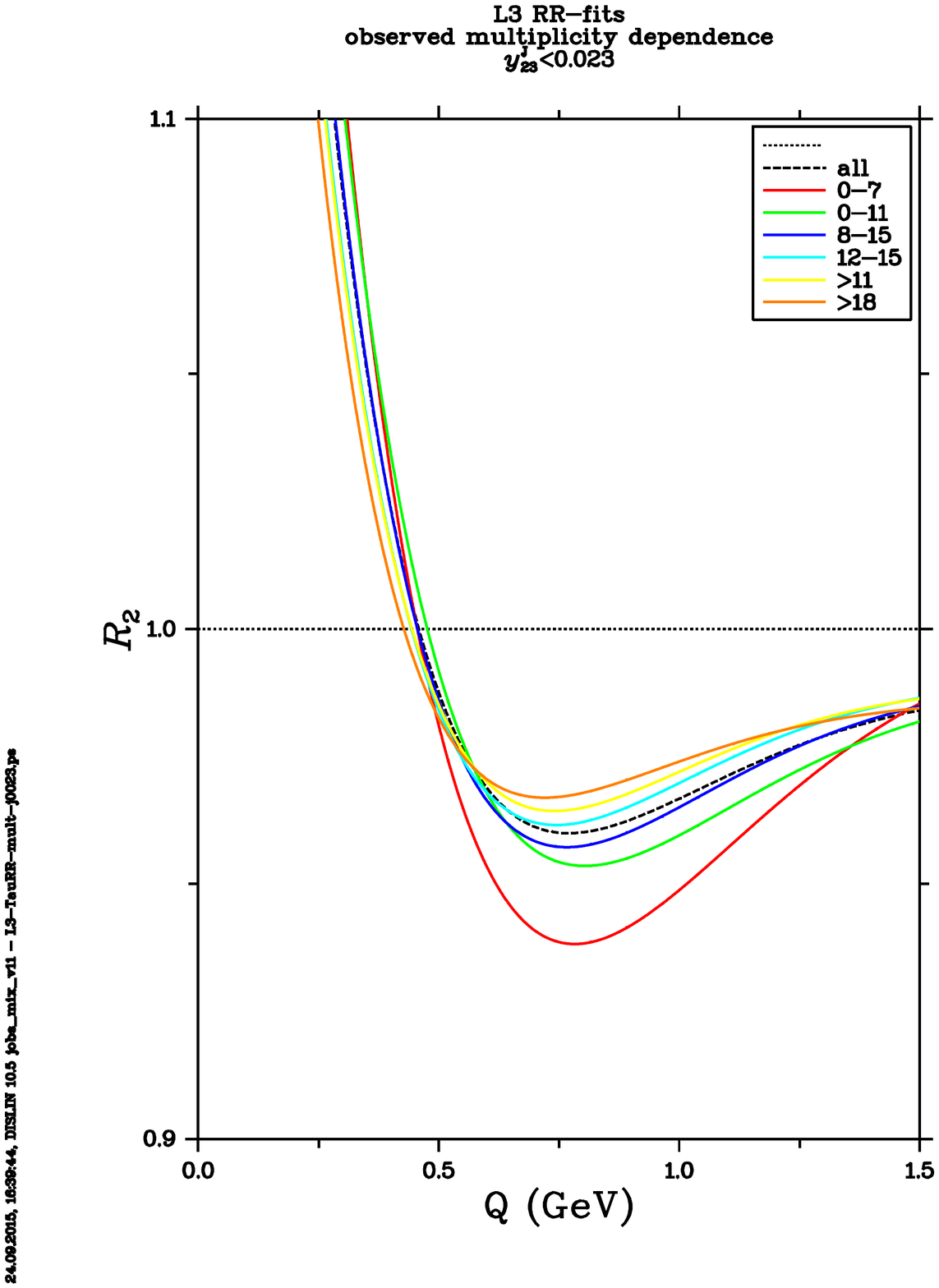}
  \hfill
  \includegraphics[width=0.43\textwidth,clip,viewport=47 59 496 653]{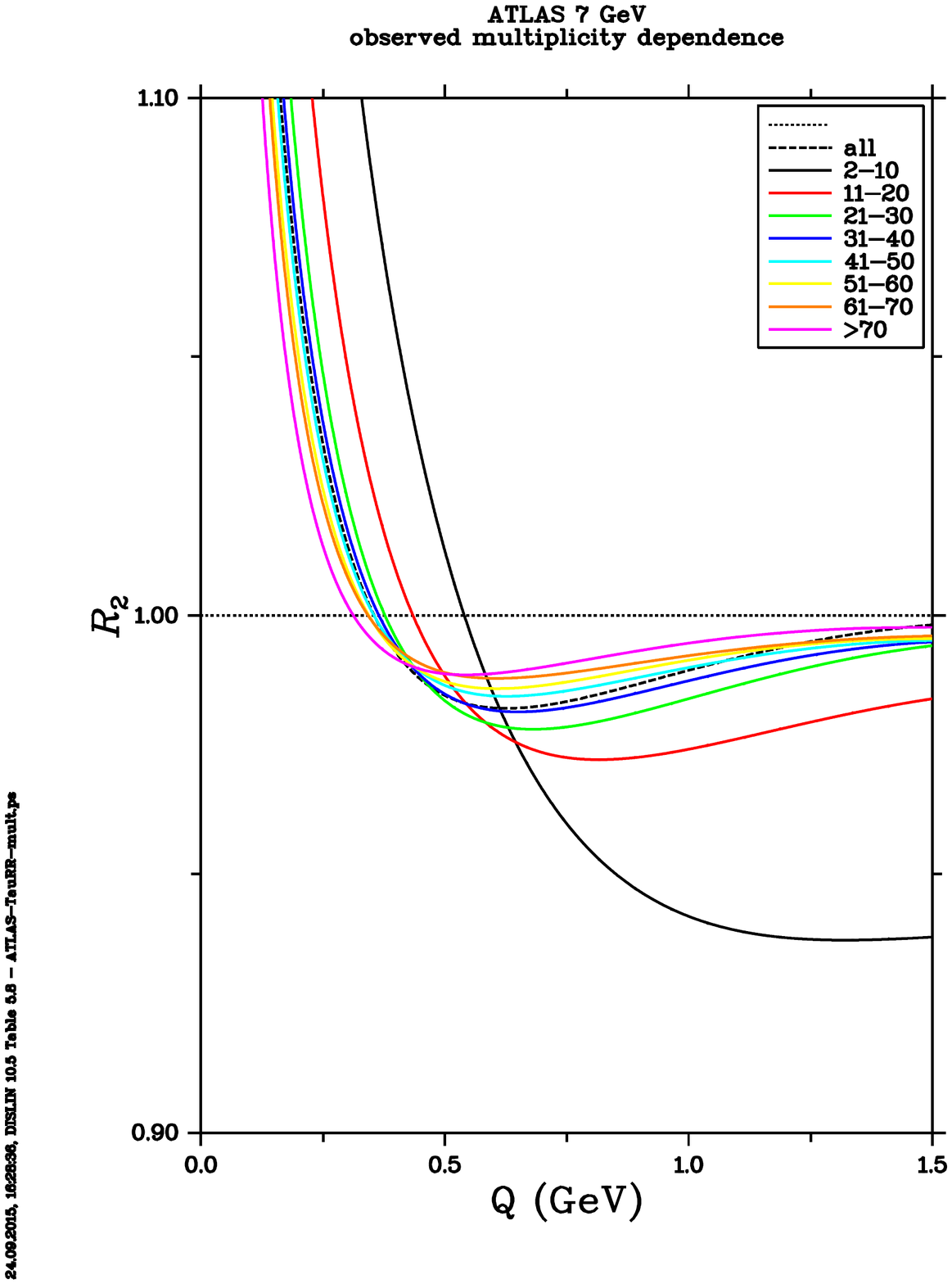}
  \caption{The result of fits of \Eq{eq:asymlevR2_} to (left) \Pee\ two-jet ($\yttJ<0.023$) and (right) pp minimum bias data
           for various intervals of multiplicity, observed in the case of \Pee\ and corrected in the case of pp.
           \label{fig:BEACmult}
          }
\end{figure*}

\Fig{fig:BEACmult} shows the dependence of the BEAC on track multiplicity.  The anti-correlation dip is deeper and at somewhat
higher $Q$ for two-jet \Pee\ than for pp.  With increasing $N$ the minimum moves to lower $Q$; this effect is larger in pp than in
\Pee.  The dip also becomes less deep as $N$ increases, an effect also noticed by \CMS~\cite{CMS:be2}.

\begin{figure*} \centering
  \includegraphics[width=0.43\textwidth,clip,viewport=47 59 496 653]{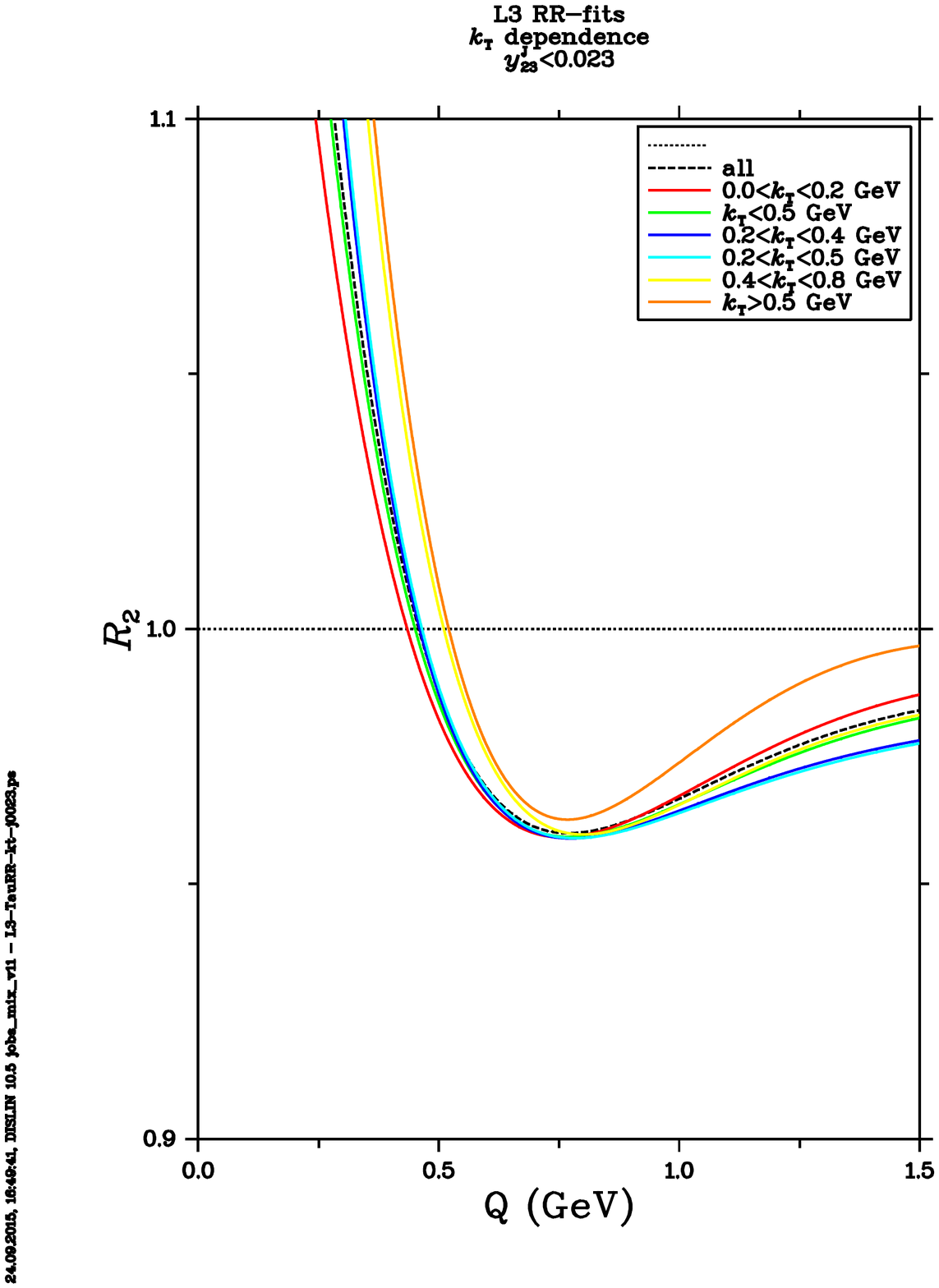}
  \hfill
  \includegraphics[width=0.43\textwidth,clip,viewport=47 59 496 653]{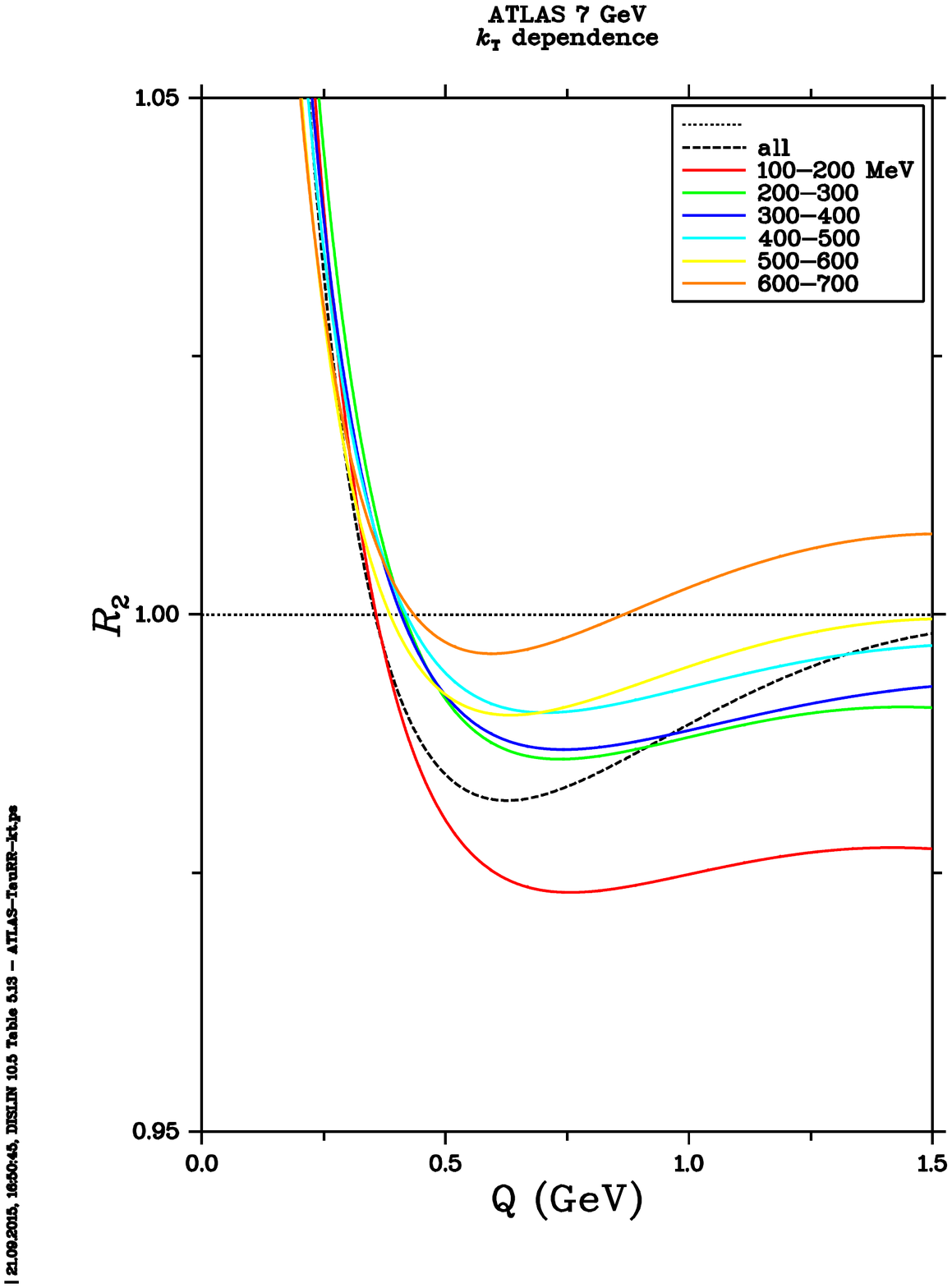}
  \caption{The result of fits of \Eq{eq:asymlevR2_} to (left) \Pee\ two-jet ($\yttJ<0.023$) and (right) pp minimum bias data
           for various intervals of \kt.
           \label{fig:BEACkt}
          }
\end{figure*}

\Fig{fig:BEACkt} shows the dependence of the BEAC on \kt. Little dependence is seen for two-jet \Pee\ data.
For pp minimum bias data, the depth of the dip decreases with \kt, but the position of the minimum is approximately constant.

\section{Conclusions}  \label{concl}
The \taumodel, which is closely related to the string picture,
provides a reasonable explanation of both the BEC peak and the BEAC dip in both \Pee\ and minimum bias pp interactions.
Of all the parametrizations tried, only that of the \taumodel\ survives as a candidate to explain the data.
Another possible explanation is that the anti-correlation arises because of the non-zero size of the pion.
To discover which of these explanations (or what combination of them) is the best explanation requires detailed investigation of
both BEC and BEAC.
 
$R$ and \Ra\ are found to depend on track multiplicity and transverse momentum in both pp and \Pee\
and on jets and rapidity in \Pee.  What more does it depend on?
If the BEAC are due to the correlation between coordinate space and momentum space, as in the \taumodel,
it is reasonable that the dependence on jet structure, as seen in \Pee, occurs.  Then one would also expect similar dependences in
pp. This possibility makes the study of BEC (and BEAC) in jet events interesting.  One might also expect a rapidity dependence
in minimum bias events, as the number of strings involved or the presence of color reconnection may vary with rapidity.
On the other hand, it is not clear (at least to me) why the effect of the pion size should depend on either rapidity or the presence
of a jet.

In studying these correlations, the reference sample used plays a crucial role.
The use of an unlike-sign reference sample is not possible in studying BEAC.
Further, the upper limit of the fit range must be well above the BEAC region.
It is very difficult, if not impossible, to compare quantitatively the results using different reference samples.
It would therefore be useful if the \LHC\ experiments could agree on a standard method of constructing the reference sample,
and give results using this standard method as well as any other method deemed superior.
This is especially important to investigate, \eg, dependence on rapidity, where the acceptances of the experiments are very
different, \eg, \LHCb\ and \ATLAS, \CMS, \ALICE.
 
\newpage
\onecolumn
\input p.bbl

\end{document}

%% file: p.bbl
\hyphenation{Post-Script Sprin-ger}